%% file: tutorial.tex
\documentclass[12pt]{article}

\usepackage{graphicx}
\usepackage{amssymb}
\usepackage{amsmath}
\usepackage{srcltx}
\usepackage{color}
\usepackage{cite}
\usepackage{url}
\usepackage{ntheorem}
\usepackage{psfrag}
\usepackage[bookmarks,colorlinks=false]{hyperref}
\usepackage{caption}
\usepackage{subcaption}

\newcommand{\mnote}[1]
{\protect{\stepcounter{mnotecount}}$^{\mbox{\footnotesize
$
\bullet$\themnotecount}}$ \marginpar{
\raggedright\tiny\em
$\!\!\!\!\!\!\,\bullet$\themnotecount: #1} }

\theorembodyfont{\upshape}

\theoremsymbol{\ensuremath{\diamondsuit}}
\newcommand{\fcoco}{\small}
\theorembodyfont{\fcoco}\theoremseparator{}\theoremindent0.5cm

\theoremstyle{nonumberplain}\theorembodyfont{\fcoco}
\theoremseparator{}\theoremindent0.5cm

\DeclareFontFamily{OT1}{rsfs}{}
\DeclareFontShape{OT1}{rsfs}{m}{n}{ <-7> rsfs5 <7-10> rsfs7 <10-> rsfs10}{}
\DeclareMathAlphabet{\mycal}{OT1}{rsfs}{m}{n}
\def\scri{{\mycal I}}%

{\catcode `\@=11 \global\let\AddToReset=\@addtoreset}
\AddToReset{equation}{section}

\newcounter{mnotecount}[section]

\AddToReset{figure}{section}
\renewcommand{\themnotecount}{\thesection.\arabic{mnotecount}}

\newcommand{\mnotex}[1]
{\protect{\stepcounter{mnotecount}}$^{\mbox{\footnotesize
$
\bullet$\themnotecount}}$ \marginpar{
\raggedright\tiny\em
$\!\!\!\!\!\!\,\bullet$\themnotecount: #1} }

\begin{document}
\title{Conformal and projection diagrams in {LaTeX}}
\date{}
\author{Christa R. \"Olz
\\
Gravitational Physics, University of Vienna
\\
\\Sebastian J. Szybka
\\
Astronomical Observatory, Jagellonian University
}
\maketitle{}
\begin{abstract}
In general relativity, the causal structure of space-time may sometimes be depicted by {\it conformal Carter-Penrose diagrams} or a recent extension of these --- the {\it projection diagrams}. The introduction of conformal diagrams in the sixties was one of the progenitors of the golden age of relativity. They are the key ingredient of many scientific papers. Unfortunately, drawing them in the form suitable for \emph{LaTeX} documents is time-consuming and not easy. We present below a library that allows one to draw an arbitrary conformal diagram in a few simple steps.
\end{abstract}
\section{Introduction}
\label{intro}
Our {\it diagrams} library is based on {\it ePiX} ---  a collection of batch utilities \cite{ePiX} for GNU/Linux and similar platforms that creates mathematically accurate figures in a form that is suitable to use with \emph{LaTeX}. The user prepares the \emph{C++} code with a description of a diagram (in a \emph{LaTeX}-like style) and then compiles the code with {\it ePiX}. The output may have the form of either a \emph{LaTeX} picture environment code, or is a \emph{pdf} or \emph{eps} file. Only very basic knowledge of {\it C++} syntax is necessary in order to be able to create a diagram.

The original motivation to write the library was the doctoral dissertation of one of us (SJS, \cite{wavemaps}), in which many conformal diagrams were presented that were constructed using a first version of the library. More recently, we co-authored a paper \cite{COS} in which an extension of the idea of a conformal diagram has been introduced. This extension was named a {\it projection diagram}. It provides a systematic procedure to visualize the four-dimensional space-time structure, as opposed to conformal diagrams which are concerned with two-dimensional space-times.
 The paper \cite{COS} contains 17 mainly nontrivial diagrams. Some more examples of diagrams constructed with the \emph{diagrams} library can be found in \cite{KdS}. In order to draw these, CR\"O extended and modified the original library.

\subsection{Dependencies, the source code and installation}

To use the library, it is necessary to have {\it LaTeX}, {\it ePiX} and {\it C++} already installed. The source of our library (\mbox{\it diagrams.h}) is available as an ancillary file to the \emph{arXiv} version of this document. This file must be copied to the path accessible by the \emph{C++} compiler (e.g.\ in linux {\it /usr/local/include}).

\subsection{Licence}

Our library was made as a scientific project, hence it is free software. Please cite this article (\emph{arXiv} reference) if you use it.

\section{Building diagrams from blocks}

In order to understand the causal properties of a space-time, one must normally find a transformation to \emph{null coordinates}. However, in many cases, symmetries simplify an analysis considerably and the causal structure may be read off directly from a form of the metric. Recently, it has been shown how to apply such a simplified analysis to four-dimensional space-times and depict their causal structure with the help of projection diagrams \cite{COS}. The procedure involves an identification of basic blocks of the diagrams and gluing them together. Conformal diagrams may also be constructed from basic blocks for several classes of space-times with the help of a simple algorithm. In most cases, the construction reduces to Walker's procedure \cite{Walker} that applies to two-dimensional Lorentzian metrics of the form
\begin{equation} \label{Krakgen2VII12}
 g=-F(r) dt^{2}+F^{-1}(r)dr^{2} \;.
\end{equation}
The structure of conformal diagrams for spherically symmetric self-similar space-times was investigated in \cite{ssss}, where the basic blocks have been identified.
\section{Construction of the diagrams}

In this chapter, the construction of diagrams is demonstrated with the example of a conformal diagram for (1+1)-dimensional Minkowski space-time. We begin by writing a \emph{C++} file that assembles the basic blocks of the diagrams and adds further elements such as labels. The procedures required for this are contained in the library \emph{diagrams.h} and in the \emph{ePiX} library \emph{epix.h}. The \emph{C++} file, let us call it \emph{minkowski.xp}, is then compiled with \emph{ePiX}, generating the file \emph{minkowski.eepic}. This can be included in a \emph{LaTeX} file, which upon compilation creates the desired diagram in \emph{dvi} or \emph{pdf} format. This file may be further converted to any desired image file format.
\subsection{\emph{C++} file}

The full \emph{C++} code is given in Appendix \ref{a1}. In this section, we provide some explanations on the code.

\subsubsection{Global structure}

The header of the file contains references to the {\it ePiX} and {\it diagrams} library and the definition of the {\it namespace}
that sets the context of the names of the procedures that we are going to use

\begin{verbatim}
  #include "epix.h"
  using namespace ePiX;
  #include "diagrams.h"
\end{verbatim}

The main function starts with a definition of the bounding box where the bottom left and the top right corners are defined. This sets our virtual canvas. The \verb+P(.,.)+ is the \emph{ePiX} point object. In order to define the real size of the picture we set units and provide the size of the picture
\newpage

\begin{verbatim}
  main()
  {
  bounding_box(P(-10,-5),P(10,11));
  unitlength("1mm");
  picture(P(80,64));

  begin();

  ...

  end();
  }
\end{verbatim}
Here ``..." stands for the main part of the example. For simplicity we split it into two parts --- diagrams and labels --- as described below.

\subsubsection{Blocks}

We start by laying down the basic structure of the diagram using the {fan} elements of the \emph{diagrams} library. By \emph{fan} we refer to an isosceles triangle, containing arrows to indicate the character of the orbits of the isometry group. For an overview of all pre-defined fans see Section \ref{bb}. Note that every fan comes with left- and with right-pointing orbits, and is designed with the longest side as base. The argument of the {fan} procedure specifies the location of the reference point (the center of its base).
A fan located at $P(0,0)$ has its base stretching from $P(-4,0)$ to $P(4,0)$, and its highest point at $P(0,4)$. Note that the length $8$ of the base is a fixed number, according to which all elements are scaled. The size of a fan is controlled in the header.

\begin{figure}
  \centering
  \begin{subfigure}[t]{0.3\textwidth}
    \centering
    \includegraphics[scale=0.7]{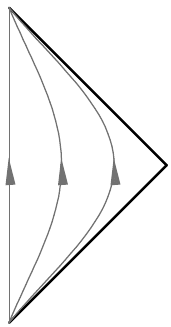}
    \caption{}
    \label{mink.a}
  \end{subfigure}
\quad
  \begin{subfigure}[t]{0.3\textwidth}
    \centering
    \includegraphics[scale=.7]{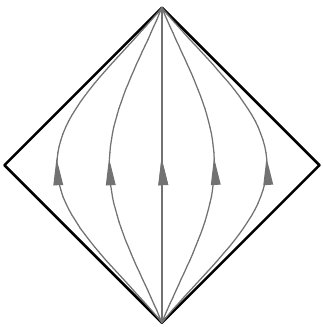}
    \caption{}
    \label{mink.b}
  \end{subfigure}
\quad
  \begin{subfigure}[t]{0.3\textwidth}
    \centering
    \includegraphics[scale=.7]{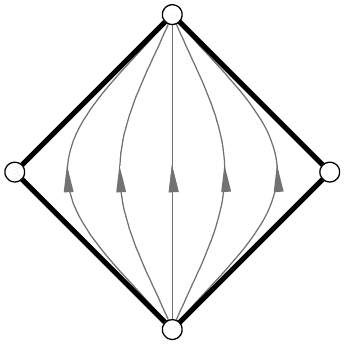}
    \caption{}
    \label{mink.c}
  \end{subfigure}
\caption{The construction of a conformal diagram for 1+1 dimensional Minkowski space-time.}
\end{figure}

\begin{figure}
  \begin{center}
  \includegraphics[scale=0.7]{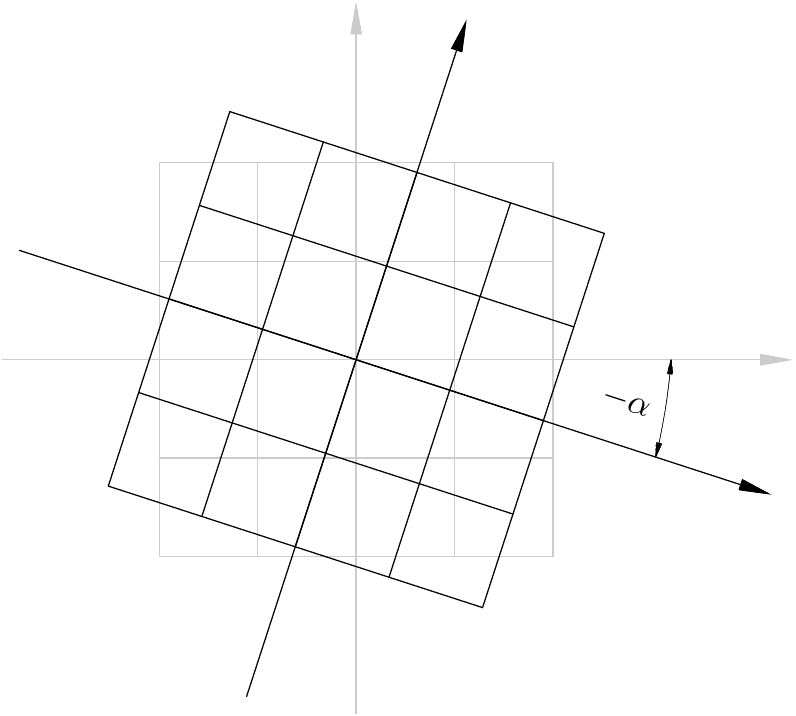}
    \caption{The effect of \texttt{camera.roll($\alpha$)} on the grid. The fixed point corresponds to $P(0,0)$.}
    \label{fig:roll}
  \end{center}
\end{figure}

For the conformal diagram of Minkowski space-time we need a fan with left-pointing orbits that is rotated clockwise by ${\pi}/{2}$ about the reference point. A fan with left-pointing orbits is indicated by the designation \emph{fan\_left} in the procedure name
and may be constructed at an arbitrary point \mbox{$P=P(0,0)$} by
\begin{verbatim}
  fan_left(P(0,0));
\end{verbatim}
To achieve the rotation, we use the procedure \emph{rotate}
\begin{verbatim}
  rotate(&fan_left,P(0,0),-pi/2);
\end{verbatim}
which creates Figure \ref{mink.a}. As a first argument of this procedure one may pass any element of the library that depends only on a single point \verb+P+\footnote{Please note the additional symbol \& before the name of a procedure which follows from the \emph{C++} syntax.}. The location \verb+P+ denotes the reference point of the element on the canvas, and the last argument corresponds to the rotation angle around \verb+P+, with positive values indicating counter-clockwise rotation.
Applying \emph{rotate} to two elements
\begin{verbatim}
  rotate(&fan_left,P(0,0),-pi/2);
  rotate(&fan_right,P(0,0),pi/2);
\end{verbatim}
we obtain Figure \ref{mink.b}, where \verb+pi+ is defined to equal the constant \verb+M_PI+ of the standard C library \emph{math.h}. Alternatively, the procedure
\begin{verbatim}
  camera.roll(double angle)
\end{verbatim}
of the {\it ePiX} library may be used to rotate elements of the library. The procedure \verb+camera.roll+ rotates the canvas, with any previously defined objects in it, counter-clockwise about a specified angle with $P(0,0)$ fixed. The effective rotation of the grid is presented in Figure \ref{fig:roll}.
Note that the canvas should be rotated back by the same angle after including the fans that we would like to appear as rotated. As an example for the use of \verb+camera.roll+, we create Figure \ref{mink.b} by using the lines
\begin{verbatim}
  camera.roll(pi/2);
    fan_left(P(0,0));
  camera.roll(pi);
    fan_right(P(0,0));
  camera.roll(pi/2);
\end{verbatim}
Additional details may be added to the diagram by addressing the relevant procedures of the \emph{diagrams} library and the {\it ePiX} library. In order to add bold lines to indicate conformal infinities, we append
\begin{verbatim}
  side_right_strong(P(0,0));
  side_left_strong(P(0,0));
  side_right_strong(P(-4,-4));
  side_left_strong(P(4,-4));
\end{verbatim}
to the previous code. Further adding circles to indicate points that are not included,
\begin{verbatim}
  out(P(4,0));
  out(P(0,4));
  out(P(-4,0));
  out(P(0,-4));
\end{verbatim}
leads to a diagram of the form as depicted in Figure \ref{mink.c}. Note that all elements are layered according to the order in which they appear in the code, with the last line adding to the top layer. This should be kept in mind when using the procedure \verb+out(P p)+, or any fans containing \verb+out(P p)+, in order to avoid having them covered by other elements.
\subsubsection{Labels}
  \label{labelssection}
To add labels to the Minkowski diagram of Figure \ref{mink.c}, we use the {label} procedure of the \emph{ePiX} library. As arguments
we provide the location of the labels on the canvas and the text content expressed in \emph{LaTeX} notation in inverted commas. Note that any backslash appearing in the \emph{LaTeX} notation must be doubled. For more information on the \emph{ePiX} procedures we refer to the \emph{ePiX} manual \cite{ePiX}. For the labels in our Minkowski diagram we use the lines
\begin{verbatim}
  label(P(2.7,2.5),"$\\scri^+$");
  label(P(-2.5,2.5),"$\\scri^+$");
  label(P(-2.5,-2.5),"$\\scri^-$");
  label(P(2.5,-2.5),"$\\scri^-$");

  label(P(0.6,4.7),"$i^+$");
  label(P(0.5,-4.6),"$i^-$");
  label(P(4.8,0),"$i^0$");
  label(P(-4.7,0),"$i^0$");
\end{verbatim}
and obtain Figure \ref{mink}. The procedure \verb+\scri+ is not defined a priori, it must be defined in the header of the \emph{LaTeX} file where the \emph{eepic} file will be included. The {label} procedure can be rotated using the \emph{ePiX} procedure \verb+label_angle(double theta)+, which rotates any subsequently specified labels counterclockwise about an angle \texttt{theta}. In order to revert to labels that are not rotated we have to specifically include
\begin{verbatim}
  label_angle(0);
\end{verbatim}
In order to label the sides of a fan, as is needed for the Schwarzschild and Kerr conformal diagrams of Sections \ref{a2}-\ref{a3}, the procedures \texttt{label1(P p, string text)} to \texttt{label4(P p, string text)} may be used. They put rotated labels to the sides of a fan, with the reference point identical to the reference point of the fan, see Figure \ref{labels} for an overview.
\subsection{\emph{LaTeX} file}
In the \emph{LaTeX} file we include \emph{minkowski.eepic}. We also define the procedure \verb+\scri+ in case it is needed in the \emph{eepic} file. The contents are then given by
\begin{verbatim}
  \documentclass{article}
  \usepackage{epic,eepic,pstricks}
  \usepackage{mathrsfs}
  \usepackage{rotating}

  \newcommand{\scri}{{\mathscr I}}

  \begin{document}
    \thispagestyle{empty}
    \input{minkowski.eepic}
  \end{document}
\end{verbatim}
\section{Elements of the library}
\label{bb}
\begin{figure}
  \centering
  \begin{subfigure}[t]{0.3\textwidth}
    \centering
    \includegraphics[scale=.7]{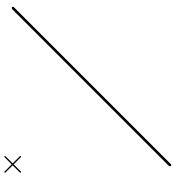}
    \caption{\texttt{side\_right(P p)}}
    \label{piece.a}
  \end{subfigure}
\quad
  \begin{subfigure}[t]{0.3\textwidth}
    \centering
    \includegraphics[scale=.7]{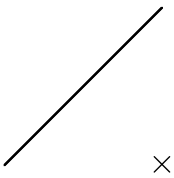}
    \caption{\texttt{side\_left(P p)}}
    \label{piece.b}
  \end{subfigure}
\quad
  \begin{subfigure}[t]{0.3\textwidth}
    \centering
    \includegraphics[scale=.7]{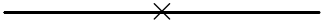}
    \caption{\texttt{base(P p)}}
    \label{piece.c}
  \end{subfigure}
\\ $\;$ \\
  \begin{subfigure}[t]{0.3\textwidth}
    \centering
    \includegraphics[scale=.7]{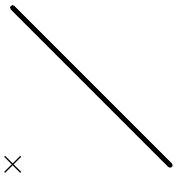}
    \caption{\texttt{side\_right\_strong(P p)}}
    \label{piece.d}
  \end{subfigure}
\quad
  \begin{subfigure}[t]{0.3\textwidth}
    \centering
    \includegraphics[scale=.7]{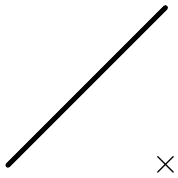}
    \caption{\texttt{side\_left\_strong(P p)}}
    \label{piece.e}
  \end{subfigure}
\quad
  \begin{subfigure}[t]{0.3\textwidth}
    \centering
    \includegraphics[scale=.7]{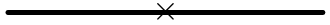}
    \caption{\texttt{base\_strong(P p)}}
    \label{piece.f}
  \end{subfigure}
\\ $\;$ \\
  \begin{subfigure}[t]{0.3\textwidth}
    \centering
    \includegraphics[scale=.7]{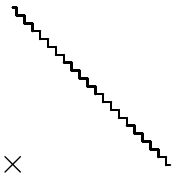}
    \caption{\texttt{side\_right\_sing(P p)}}
    \label{piece.g}
  \end{subfigure}
\quad
  \begin{subfigure}[t]{0.3\textwidth}
    \centering
    \includegraphics[scale=.7]{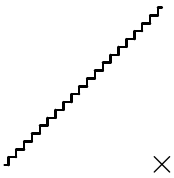}
    \caption{\texttt{side\_left\_sing(P p)}}
    \label{piece.h}
  \end{subfigure}
\quad
  \begin{subfigure}[t]{0.3\textwidth}
    \centering
    \includegraphics[scale=.7]{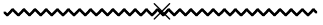}
    \caption{\texttt{base\_sing(P p)}}
    \label{piece.i}
  \end{subfigure}
\\ $\;$ \\ $\;$ \\
  \begin{subfigure}[t]{0.3\textwidth}
    \centering
    \includegraphics[scale=.7]{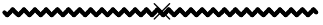}
    \caption{\texttt{base\_sing\_strong(P p)}}
    \label{piece.j}
  \end{subfigure}
\quad
  \begin{subfigure}[t]{0.3\textwidth}
    \centering
    \includegraphics[scale=.7]{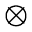}
    \caption{\texttt{out(P p)}}
    \label{piece.m}
  \end{subfigure}
\quad
  \begin{subfigure}[t]{0.3\textwidth}
    \centering
    \includegraphics[scale=.7]{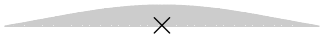}
    \caption{\texttt{takeout(P p, {double s}, {double c})%
    }}
    \label{piece.n}
  \end{subfigure}
\\ $\;$ \\
  \begin{subfigure}[t]{0.3\textwidth}
    \centering
    \includegraphics[scale=.7]{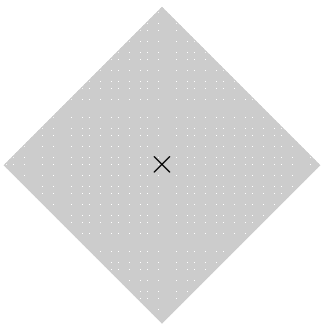}
    \caption{\texttt{diamond\_fill(P p)%
    }}
    \label{piece.o}
  \end{subfigure}
\quad
  \begin{subfigure}[t]{0.3\textwidth}
    \centering
    \includegraphics[scale=.7]{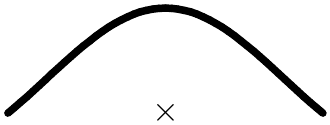}
    \caption{\texttt{curve\_w(P p, {double s}, {double c}, {double w})%
    }}
    \label{piece.k}
  \end{subfigure}
\quad
  \begin{subfigure}[t]{0.3\textwidth}
    \centering
    \includegraphics[scale=.7]{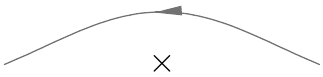}
    \caption{\texttt{curve(P p, {double s}, {double c}) and \\
     arrowhead(P p, {double s}, {int dir})%
    }}
    \label{piece.l}
  \end{subfigure}
\caption{The elements of the \emph{diagrams} library used to create the fans. The parameter \texttt{s} defines the curvature, from a straight line (\texttt{s=0}) approaching a triangle for large \texttt{s}, the parameter \texttt{c} specifies the shading from \texttt{c=0} (white) to \texttt{c=1} (black), \texttt{w} gives the line width and \texttt{dir} defines the direction of the arrow, with \texttt{+1} for left-pointing and \texttt{-1} for right-pointing.}
\label{pieces}
\end{figure}

\begin{figure}
  \centering
  \begin{subfigure}[t]{0.47\textwidth}
    \centering
    \includegraphics[scale=.7]{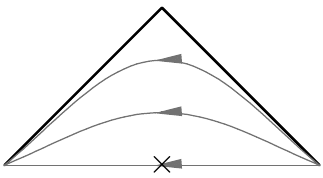}
    \caption{\texttt{fan\_left(P p)}}
  \end{subfigure}
\quad
  \begin{subfigure}[t]{0.47\textwidth}
    \centering
    \includegraphics[scale=.7]{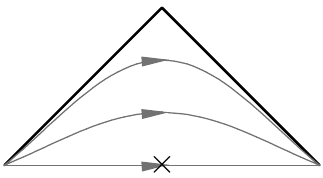}
    \caption{\texttt{fan\_right(P p)}}
    \label{fans.b}
  \end{subfigure}
\\ $\;$ \\
  \begin{subfigure}[t]{0.47\textwidth}
    \centering
    \includegraphics[scale=.7]{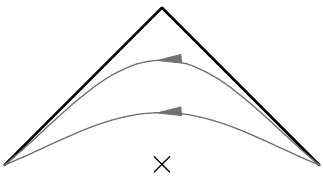}
    \caption{\texttt{fan\_left\_empty(P p)}}
    \label{fans.c}
  \end{subfigure}
\quad
  \begin{subfigure}[t]{0.47\textwidth}
    \centering
    \includegraphics[scale=.7]{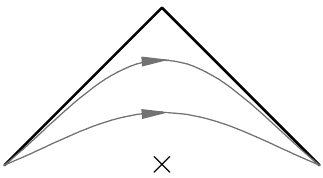}
    \caption{\texttt{fan\_right\_empty(P p)}}
    \label{fans.d}
  \end{subfigure}
\\ $\;$ \\
  \begin{subfigure}[t]{0.47\textwidth}
    \centering
    \includegraphics[scale=.7]{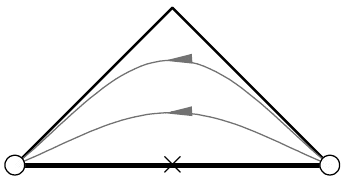}
    \caption{\texttt{fan\_left\_strong(P p)}}
    \label{fans.e}
  \end{subfigure}
\quad
  \begin{subfigure}[t]{0.47\textwidth}
    \centering
    \includegraphics[scale=.7]{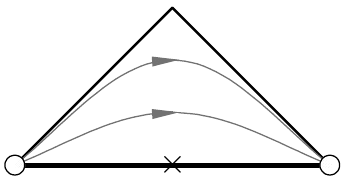}
    \caption{\texttt{fan\_right\_strong(P p)}}
    \label{fans.f}
  \end{subfigure}
\\ $\;$ \\
  \begin{subfigure}[t]{0.47\textwidth}
    \centering
    \includegraphics[scale=.7]{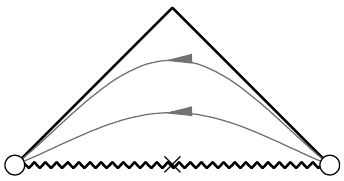}
    \caption{\texttt{fan\_left\_sing(P p)}}
    \label{fans.g}
  \end{subfigure}
\quad
  \begin{subfigure}[t]{0.47\textwidth}
    \centering
    \includegraphics[scale=.7]{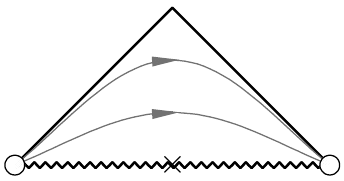}
    \caption{\texttt{fan\_right\_sing(P p)}}
    \label{fans.h}
  \end{subfigure}
\\ $\;$ \\
  \begin{subfigure}[t]{0.47\textwidth}
    \centering
    \includegraphics[scale=.7]{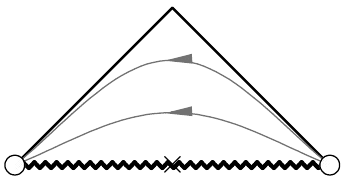}
    \caption{\texttt{fan\_left\_sing\_strong(P p)}}
    \label{fans.i}
  \end{subfigure}
\quad
  \begin{subfigure}[t]{0.47\textwidth}
    \centering
    \includegraphics[scale=.7]{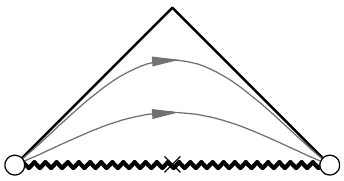}
    \caption{\texttt{fan\_right\_sing\_strong(P p)}}
    \label{fans.j}
  \end{subfigure}
\\ $\;$ \\
  \begin{subfigure}[t]{0.47\textwidth}
    \centering
    \includegraphics[scale=.7]{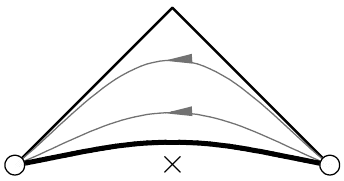}
    \caption{\texttt{fan\_left\_curved\_strong(P p)}}
    \label{fans.q}
  \end{subfigure}
\quad
  \begin{subfigure}[t]{0.47\textwidth}
    \centering
    \includegraphics[scale=.7]{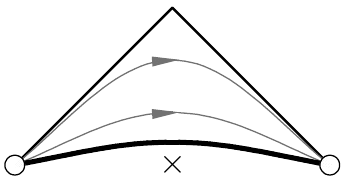}
    \caption{\texttt{fan\_right\_curved\_strong(P p)}}
    \label{fans.r}
  \end{subfigure}
\\ $\;$ \\
  \begin{subfigure}[t]{0.47\textwidth}
    \centering
    \includegraphics[scale=.7]{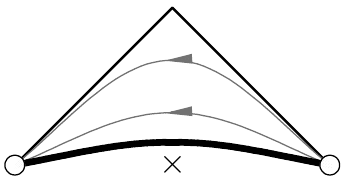}
    \caption{\texttt{fan\_left\_curved\_very\_strong(P p)}}
    \label{fans.s}
  \end{subfigure}
\quad
  \begin{subfigure}[t]{0.47\textwidth}
    \centering
    \includegraphics[scale=.7]{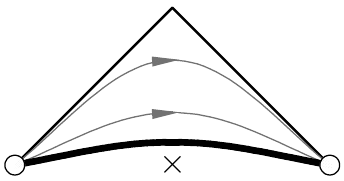}
    \caption{\texttt{fan\_right\_curved\_very\_strong(P p)}}
    \label{fans.t}
  \end{subfigure}
\caption{An overview of all fans defined in the \emph{diagrams} library without shaded area.}
\label{fans}
\end{figure}

\begin{figure}
  \centering
    \begin{subfigure}[t]{0.47\textwidth}
    \centering
    \includegraphics[scale=.7]{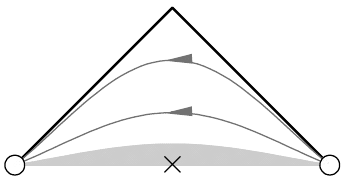}
    \caption{\texttt{fan\_left\_takeout(P p)}}
    \label{fans.k}
  \end{subfigure}
\quad
  \begin{subfigure}[t]{0.47\textwidth}
    \centering
    \includegraphics[scale=.7]{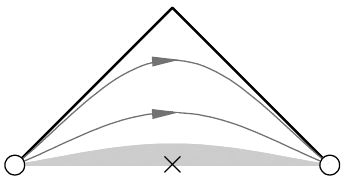}
    \caption{\texttt{fan\_right\_takeout(P p)}}
    \label{fans.l}
  \end{subfigure}
\\ $\;$ \\
  \begin{subfigure}[t]{0.47\textwidth}
    \centering
    \includegraphics[scale=.7]{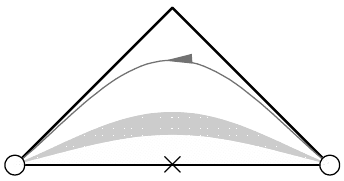}
    \caption{\texttt{fan\_left\_takeout\_curved(P p)}}
    \label{fans.m}
  \end{subfigure}
\quad
  \begin{subfigure}[t]{0.47\textwidth}
    \centering
    \includegraphics[scale=.7]{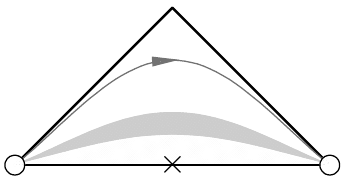}
    \caption{\texttt{fan\_right\_takeout\_curved(P p)}}
    \label{fans.n}
  \end{subfigure}
\\ $\;$ \\
  \begin{subfigure}[t]{0.48\textwidth}
    \centering
    \includegraphics[scale=.7]{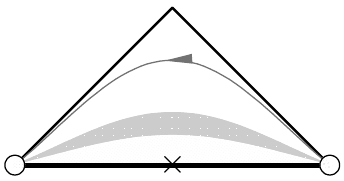}
    \caption{\texttt{fan\_left\_takeout\_curved\_strong(P p)}}
    \label{fans.o}
  \end{subfigure}
\quad
  \begin{subfigure}[t]{0.48\textwidth}
    \centering
    \includegraphics[scale=.7]{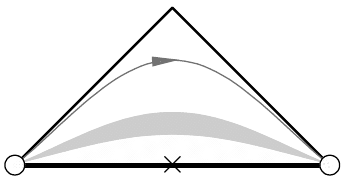}
    \caption{\texttt{fan\_right\_takeout\_curved\_strong(P p)}}
    \label{fans.p}
  \end{subfigure}
\caption{An overview of all fans defined in the \emph{diagrams} library containing a shaded area.}
\label{fans2}
\end{figure}

\begin{figure}
\vspace{-0.5cm}
  \centering
    \begin{subfigure}[t]{\textwidth}
    \centering
    \includegraphics[scale=.7]{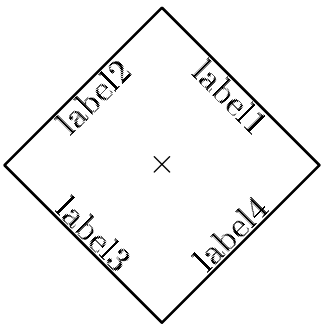}
    \end{subfigure}
\vspace{0.5cm}
  \caption{The procedures \texttt{label1(P p, string text)}, \texttt{label2(P p, string text)}, \texttt{label3(P p, string text)} and \texttt{label4(P p, string text)}, with the side lines indicating the shape of a fan and its mirror image located at $\times$.}
  \label{labels}
\end{figure}

This section gives an overview of the procedures defined in the diagrams library and their dependence on the parameters, see Figures \ref{pieces}-\ref{labels}. The reference point addressed by the parameter \texttt{p} is indicated by the symbol $\times$ in the figures.
\par \vspace{4mm}
\captionsetup[subfigure]{labelformat=empty}

%
%
\clearpage

\bigskip

\noindent{\sc Acknowledgements.}
We thank Piotr Chru\'sciel, Micha\l~Eckstein and James Grant for useful comments.
\appendix
\section{Sample files}
\begin{figure}
\begin{center}
{\includegraphics[scale=.7]{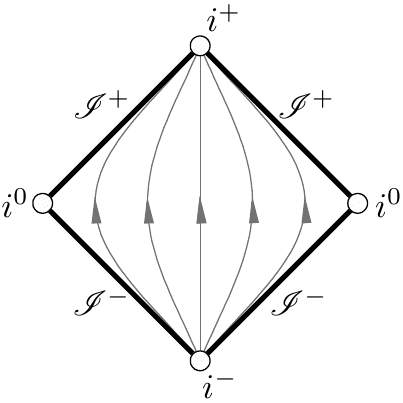}}
\caption{A conformal diagram of Minkowski space-time.}
\label{mink}
\end{center}
\end{figure}

\begin{figure}
\centering
{\includegraphics[scale=.7]{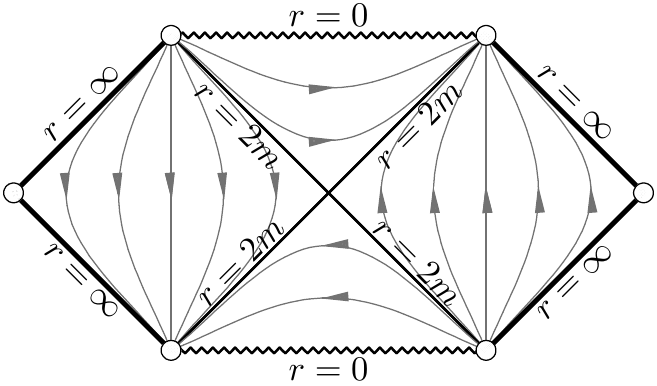}}
\caption{A conformal diagram of Schwarzschild space-time.
\label{schw}}
\end{figure}

\begin{figure}
\centering
{\includegraphics[scale=.7]{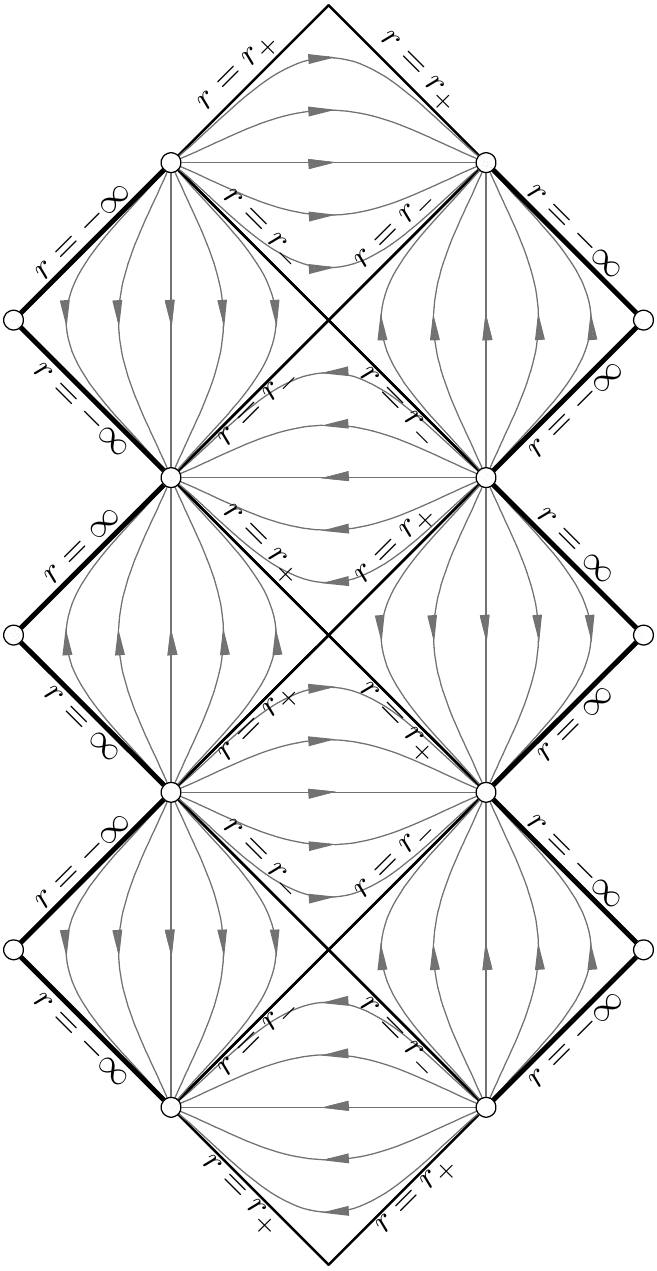}}
\caption{A conformal diagram for the non-extremal Kerr space-time at the symmetry axis, with horizons at $r_{\pm}$.
\label{kerr}}
\end{figure}

Three sample files are given, used to create the conformal diagrams for Minkowski, Schwarzschild and the non-extremal Kerr space-time on the symmetry axis. Figures (\ref{mink})-(\ref{kerr}) show the resulting diagrams.
\subsection{Minkowski}\label{a1}
\input{mink_sample.tex}

\subsection{Schwarzschild} \label{a2}
\input{schwarzschild_sample.tex}
\subsection{Non-extremal Kerr at the axis of symmetry} \label{a3}
\input{kerr_sample.tex}

\bibliographystyle{amsplain}
\bibliography{tutorial}
\end{document}

%% file: mink_sample.tex
\begin{verbatim}
  #include "epix.h"
  using namespace ePiX;
  #include "diagrams.h"

  main()
  {
  bounding_box(P(-10,-5),P(10,11));
  unitlength("1mm");
  picture(P(80,64));

  begin();

  rotate(&fan_left,P(0,0),-pi/2);
  rotate(&fan_right,P(0,0),pi/2);

  side_right_strong(P(0,0));
  side_left_strong(P(0,0));
  side_right_strong(P(-4,-4));
  side_left_strong(P(4,-4));

  out(P(4,0));
  out(P(0,4));
  out(P(-4,0));
  out(P(0,-4));

  label(P(2.7,2.5),"$\\scri^+$");
  label(P(-2.5,2.5),"$\\scri^+$");
  label(P(-2.5,-2.5),"$\\scri^-$");
  label(P(2.5,-2.5),"$\\scri^-$");

  label(P(0.6,4.7),"$i^+$");
  label(P(0.5,-4.6),"$i^-$");
  label(P(4.8,0),"$i^0$");
  label(P(-4.7,0),"$i^0$");

  end();
  }
\end{verbatim}

%% file: schwarzschild_sample.tex
\begin{verbatim}
  #include "epix.h"
  using namespace ePiX;
  #include "diagrams.h"

  main()
  {
  bounding_box(P(-10,-5),P(10,11));
  unitlength("1mm");
  picture(P(80,64));

  begin();

  rotate(&fan_left,P(-4,0),pi/2);
  rotate(&fan_right,P(-4,0),-pi/2);
  rotate(&fan_right,P(4,0),pi/2);
  rotate(&fan_left,P(4,0),-pi/2);

  side_left_strong(P(-4,0));
  side_right_strong(P(-8,-4));
  side_right_strong(P(4,0));
  side_left_strong(P(8,-4));

  fan_left_sing(P(0,-4));
  rotate(&fan_left_sing,P(0,4),pi);

  out(P(8,0));
  out(P(-8,0));

  label(P(0,4.5),"$r=0$");
  label(P(0,-4.5),"$r=0$");

  label2(P(4,0),"$r=2m$");
  label3(P(4,0),"$r=2m$");
  label1(P(-4,0),"$r=2m$");
  label4(P(-4,0),"$r=2m$");

  label3(P(8,4),"$r=\\infty$");
  label2(P(8,-4),"$r=\\infty$");
  label4(P(-8,4),"$r=\\infty$");
  label1(P(-8,-4),"$r=\\infty$");

  end();
  }

\end{verbatim}

%% file: kerr_sample.tex
\begin{verbatim}
  #include "epix.h"
  using namespace ePiX;
  #include "diagrams.h"

  main()
  {
  bounding_box(P(-10,-5),P(10,11));
  unitlength("1mm");
  picture(P(80,64));

  begin();

  fan_left(P(0,0));
  fan_right(P(0,8));
  fan_left(P(0,-16));
  fan_right(P(0,-8));

  rotate(&fan_right,P(0,0),pi);
  rotate(&fan_left,P(0,8),pi);
  rotate(&fan_right,P(0,-16),pi);
  rotate(&fan_left,P(0,-8),pi);

  rotate(&fan_left,P(4,4),-pi/2);
  rotate(&fan_right,P(4,-4),-pi/2);
  rotate(&fan_left,P(4,-12),-pi/2);

  rotate(&fan_left,P(-4,4),pi/2);
  rotate(&fan_right,P(-4,-4),pi/2);
  rotate(&fan_left,P(-4,-12),pi/2);

  rotate(&fan_right,P(4,4),pi/2);
  rotate(&fan_left,P(4,-4),pi/2);
  rotate(&fan_right,P(4,-12),pi/2);

  rotate(&fan_right,P(-4,4),-pi/2);
  rotate(&fan_left,P(-4,-4),-pi/2);
  rotate(&fan_right,P(-4,-12),-pi/2);

  side_left_strong(P(8,0));
  side_right_strong(P(-8,0));

  side_left_strong(P(8,-8));
  side_right_strong(P(-8,-8));

  side_left_strong(P(8,-16));
  side_right_strong(P(-8,-16));

  side_left_strong(P(-4,4));
  side_right_strong(P(4,4));

  side_left_strong(P(-4,-4));
  side_right_strong(P(4,-4));

  side_left_strong(P(-4,-12));
  side_right_strong(P(4,-12));

  out(P(8.0,4));			
  out(P(-8.0,4));
  out(P(8,-4));
  out(P(-8,-4));
  out(P(8.0,-12));
  out(P(-8.0,-12));

  out(P(4,0));
  out(P(-4,0));
  out(P(4,8));
  out(P(-4,8));
  out(P(4,-8));
  out(P(-4,-8));
  out(P(4,-16));
  out(P(-4,-16));

  label1(P(0,0),"$r=r_-$");
  label2(P(0,0),"$r=r_-$");
  label3(P(0,0),"$r=r_+$");
  label4(P(0,0),"$r=r_+$");
  label3(P(0,8),"$r=r_-$");
  label4(P(0,8),"$r=r_-$");
  label1(P(0,-8),"$r=r_+$");
  label2(P(0,-8),"$r=r_+$");
  label3(P(0,-8),"$r=r_-$");
  label4(P(0,-8),"$r=r_-$");
  label1(P(0,-16),"$r=r_-$");
  label2(P(0,-16),"$r=r_-$");

  label4(P(-4,12),"$r=r_+$");
  label3(P(4,12),"$r=r_+$");
  label1(P(-4,-20),"$r=r_+$");
  label2(P(4,-20),"$r=r_+$");

  label4(P(-8,8),"$r=-\\infty$");
  label1(P(-8,0),"$r=-\\infty$");
  label4(P(-8,0),"$r=\\infty$");
  label1(P(-8,-8),"$r=\\infty$");
  label4(P(-8,-8),"$r=-\\infty$");
  label1(P(-8,-16),"$r=-\\infty$");

  label3(P(8,8),"$r=-\\infty$");
  label2(P(8,0),"$r=-\\infty$");
  label3(P(8,0),"$r=\\infty$");
  label2(P(8,-8),"$r=\\infty$");
  label3(P(8,-8),"$r=-\\infty$");
  label2(P(8,-16),"$r=-\\infty$"); 

  end();
  }
\end{verbatim}